# Yusuf Maitama Sule University, Kano
## Faculty of Science
### 3rd Annual International Conference

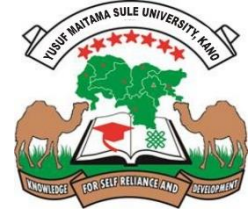

# Design and Comparison Migration between Ipv4 and Ipv6 Transition Techniques

**Abubakar Isa and Idris Abdulmumin**
Department Computer Science, Ahmadu Bello University Zaria, Kaduna, Nigeria
**Email**: sadiqrrw@gmail.com; abumafrim@gmail.com

**ABSTRACT**
IPv4 which is the old version of Internet Protocol has a new successor named IP Next Generation (IPng) or IPv6 developed by Internet Engineering Task Force (IETF). This new version is developed specifically to resolve some issues of IPv4 like scalability, performance and reliability. Although new version is ready for usage, it is obvious that it will take years to transit fully from IPv4 to IPv6. We have to use these two versions together for a long time. Therefore, we have to investigate and know transition mechanisms that we can use during transition period to achieve a transition with minimum problem. This research defines the essential information about compatibility transition mechanisms between IPv4-IPv6. Dual Stack is one of the IPv4-IPv6 compatible mechanism by running both IPv4 stack and IPv6 stack in a single node. 6 to 4 tunneling mechanism encrypts IPv6 packets in IPv4 packets to make communications possible, from IPv6 network over IPv4 network. This has been configured using GNS3 Simulator and Packet Tracer 6.1. Dual Stack & Tunneling mechanisms were completely implemented later in this research work. This research examines transmission latency, throughput and delay from end to end, through empirical observations of both Dual Stack and tunneling mechanisms.

## I. INTRODUCTION

In today's information age, the increasing use of the Internet in the last two decades has shown the potential that the Internet can change and improve different areas such as education, businesses or entertainment, etc. No one could guess that World Wide Web would become a worldwide communication channel, and it was thought that the number of addresses provided by the Internet Protocol version 4 (IPv4) were more than enough. IPv4 developed in 1981 and was used to make interconnection between different networks. After three versions, IP got the name of next version number 4 and declared as IPv4 to the internet community. The first version of IPv4 was generally used to ensure that two computers or any two network devices could connect with each other. As there is an ever growing expansion and advancement in the network and internet mechanism, the requirement of unique addresses is increasing. Therefore, to solve the address limitation of current IPv4, several technologies have come out like Network Address Translation (NAT) and Dynamic Host Configuration Protocol (DHCP). However, while these technologies decrease the addressing shortage, they prevent IP level end-to-end security, reduce robustness, so they are not good solutions [1].

To fix the problem of current IPv4, a new internet protocol was developed, which is called Internet Protocol version 6 (IPv6). This protocol was developed by the Internet Engineering Task Force (IETF) with reference to routing addresses and security. IPv6, actually is known as IP next generation (IPng), is chosen from numerous suggested alternatives as the most appropriate successor of the present Internet Protocol (IPv4). IPv6 is more effective, scalable, secure and routable than IPv4.

The reason to create a new Internet Protocol (IPv6) is basically to boost the quantity of IP address space [2]. The IPv6 can give above $3.4 \times 10^{38}$ unique addresses as compared to IPv4 which gives $4.3 \times 10^{9}$ unique addresses (IPv6 has the capacity of 128-bit/16 bytes' address scheme, whereas IPv4 has just 32 bits/4 bytes). This means, IPv6 solves the problem by eradicating the requirement of Network Address. It easily provides all devices like MP3 player, telephone, mobile phone or automobiles their own IP addresses. Moreover, it also supports multimedia transmissions, security and scalability. This proves that IPv6 was modeled by keeping in consideration the future applications. Therefore, numerous organizations like Department of Defense (DoD) of USA, have made a timetable to implement the new IPv6 for their requirement of future deployments [3]

### I. Internet Protocol (IPV4)

The network-layer protocol, i.e. Internet Protocol (IP), possesses not only control information but also addressing information which makes packets routed. IP, as the major network layer protocol in the internet protocol suite, is documented in RFC 791. There are two main duties of IP; the first one is to provide connectionless datagrams best effort delivery, and the second one is to support maximum transmission unit (MTU) sized fragmentation and reassembly of datagrams [5].

Today almost 40% of the people in the world use internet connection. Whereas, it was not more than 1% in 1995, in fact it was even less. As seen in Figure 2.1, there is an enormous increase of internet users from the year 1999 to 2013 which is ten times more. The number of Internet users reached to 1 billion in 2005, 2 billion in 2010, and is around 3 billion when we come to 2016 [6].

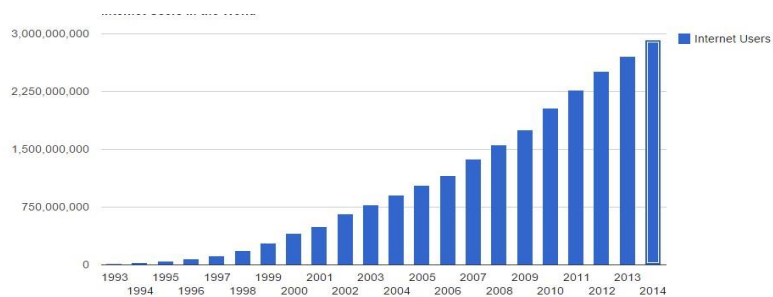

**Figure 1: Internet Users in the World**

### I. New Version of IP (IPV6)

The Internet Engineering Task Force (IETF) came up with a solution called IP next generation (IPng) to solve the IP address exhaustion problem. IPng was a result of the proposals reviewed by IETF. IPng was not a complete protocol, but was a product which had to be reviewed considering the features and limitations. After multiple reviews and changes to IPng, IPv6 was developed [10] [11].

The IPv6 address size is 128 bits compared to the 32-bit address in IPv4. The 128-bit size gives approximately 1500 addresses per square foot of the earth's surface. Even if every device around you is IP capable, there are 1500 addresses per square foot which is sufficient for any kind of requirements. Thus, IPv6 has provided a solution to IP address exhaustion. The Internet community is taking time to adapt to IPv6. The main reason would be that it is difficult for IPv4 and IPv6 to coexist. Whenever an IPv6 host wants to communicate with an IPv4 host, it has to use transmission mechanisms. It may be predicted that when IPv4 addresses are exhausted the Internet community will be forced to adopt IPv6 conversion at a faster rate.

### II. Benefits and Characteristics of IPV6 Usage

Apart from fulfilling the expected demands for the future address requirements, the advantages of using IPv6 for the skilled IT people are given below:

**Scalability:** IPv6 possesses 128-bit addresses compared to 32-bit IPv4 addresses. IPv6 provides $2^{128}$ theoretical addresses versus $2^{32}$ addresses of IPv4 [8].

**Auto configuration and "Plug-and-Play":** Plug and Play technology enables IPv6 devices to configure them independently. The device determines its address which will probably be unique based on the network prefix and its Ethernet MAC address. With this support, it is possible to plug a node into an IPv6 network without requiring any human intervention. This support is very important for the new mobile systems and its various services. As a result, network devices could connect to the network without manual configuration and without any servers such as Dynamic Host Configuration Protocol (DHCP) servers.

**Mobility:** Contrary to the Mobile IPv4 protocol, the Mobile IPv6 (MIPv6) helps avoid triangular routing experienced earlier, and makes it possible for mobile (WiFi) clients to select a new router without renumbering, which results in a more reliable and faster connection with less network interruption.

**Security:** IPv6 protocol has been developed so that the security features become an integral part of the protocol to ensure maximum security. Added security features do not affect the performance, speed and efficiency. Security features that have been added to the IPv4 protocol later as something optional are now found as a part of protocol implicitly in IPv6. IPv6 encrypts data and examines the integrity of the packets similar to those offered by the VPN data transmitted over the Internet [9].

**Quality of Service:** The quality of service in IPv6 can be handled in the same way as IPv4. Traffic Class support of IPv6 works well with the Differentiated Service model of Internet Engineering Task Force (IETF). In addition, the header of IPv6 has a new field called flow label, which can contain a label specifying a particular flow such as video or video stream. The source node generates this flow label. The existence of flow label enables devices on the way to take appropriate action based on this marker for quality of service.

### III. Comparison of IPV4 and IPV6 Header

The format of IPv6 header is simpler in comparison to IPv4, even though the quantity of address of IPv6 makes its header increase in size. The header size of IPv4 is basically just 20 octets; however, the options field variable length further builds the total IPv4 packet size. The header of IPv6 has a size of 40 octets. 6 IPv4 header fields are extracted out of 12 in IPv6. Some of the fields of the former protocol have been taken over by changing as well as adapting names. Some new fields are added to infuse new features as can be seen in Figure 2.3 [7].

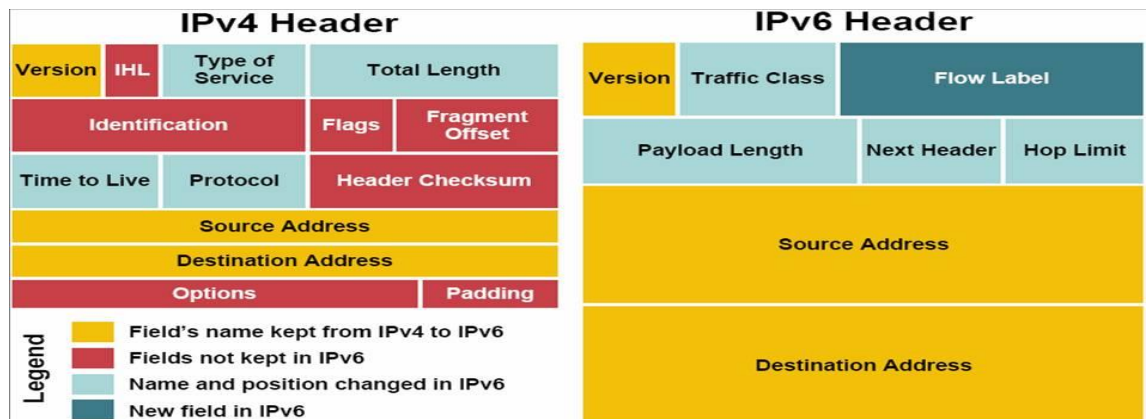

**Figure 0: IPv4 and IPv6 Header Comparison**

Even though the removal helps simpler IPv6 header to process fast, whereas, overall performance as well as the routing efficiency is dependent on the treatment of option headers as well as the algorithms that any given device should run. Apart from this, IPv6 has the advantage of having 64-bit processors of the present generation, because of which IPv6 header fields comprise of 64 bits.

### 2. Transition Mechanisms
### 2.1 Introduction

The deployment of IPv6 is happening gradually on the Internet. Initially deployment of IPv6 should be happened within isolated islands, and then these islands should interconnect with other islands over existing IPv4 Infrastructures. To connect these islands "Transition mechanisms" are needed and these are compatible mechanisms employed for Co-existence of IPv6 and IPv4 infrastructure.

A number of transition mechanisms have been defined to support co-existence [12]. There is an additional need for a node to support IPv6. RFC (request for comments) defines the following types of nodes [12][RFC 4294].

-       A node is device that has
IP implements in it.

-       A router is a device that forwards IP packets towards destination.

-       A host is a node but not a
router.

**Types of node:**
- **IPv4-only node:** A host or router that implements only IPv4 protocol in it.
- **IPv6/IPv4 node:** A host or router that implements both IPv6 and IPv4 protocol in it.
- **IPv6-only node:** A host or router that implements only IPv6 protocol in it.

This research focuses on the transition mechanisms from IPv4 to IPv6 networks, in particular the Dual-stack, Automatic 6to4 and Manual 6in4. This research provides the detailed information for the transition mechanisms. Figure 3.1 describes the transition mechanisms from IPv4 to IPv6, while Figure 3.2 elaborates the Dual Stack Transition Mechanism. Meanwhile, Figure 3.3 explains the tunneling mechanism. Figure 3.4 explains the translation mechanism.

**2.2    Transition Mechanisms from IPv4 to IPv6**

The IPv6 which is the latest version of internet protocol is not backward compatible with IPv4, which means that IPv6 networks cannot communicate with IPv4 networks. That is, they cannot send packets to each other, IPv6 network can only send IPv6 packets to other IPv6 networks, as well as the IPv4 network can send to other IPv4 networks. Therefore, there is an important interoperability problem with the coexistence of both protocols on the internet [12].

In order to solve the communication problem between IPv4 network and IPv6 network and make packets transmission between networks smoothly, the Internet Engineering Task Force (IETF) and Next Generation Transition (NGtrans) work group established IPv4/IPv6 transition mechanisms to eliminate lack of compatibility problem and support co-existence of protocols, which will last probably for a very long time. To comprehend the transition mechanisms and their importance, it is necessary to study and analyze each transition mechanism properly. The transition mechanisms are divided into three main groups as dual-stack, tunnels (including configured and automatic tunnels), and translation mechanisms. These mechanisms are explained one by one in detail below.

**2.3    Tunnels**

Tunneling is a process of encapsulating one protocol into other protocol. Wrapping an IPv6 packet within an IPv4 packet is shown in Figure 3.3. With the help of such tunneling mechanism, a packet can be carried by an incompatible network towards the destination network [13]. The tunneling migration strategy is that, a node encapsulates an IPv6 packet in an IPv4 packet for transmission across an IPv4 network and then the packet is de-capsulated to the original IPv6 packet by another node. The tunnel mechanism can be utilized to establish connection between IPv6 networks which are in isolation. Unfortunately, this is not a permanent solution. When dual-stack or native IPv6 is fully implemented, there will be no requirement for tunneling strategies [4] [13].

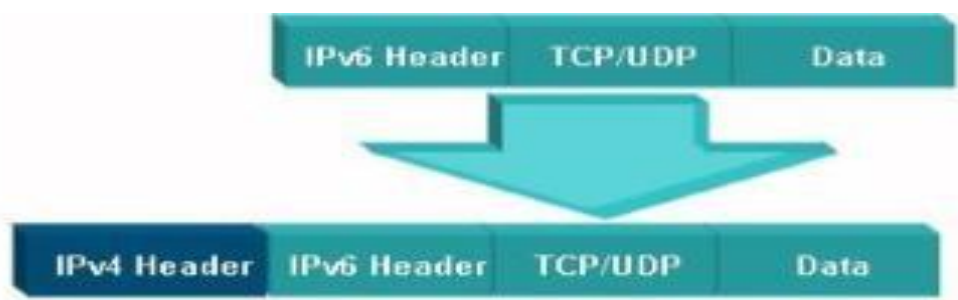

**Figure 3: IPv6 over IPv4 Tunneling**

**2.4    Configured Tunneling (Manual Tunneling)**

If network administrators manually configure the tunnel within the end devices, this is called 'configured tunneling' or 'explicit tunneling'. Configuration information that is stored in encapsulating end devices is used to determine the addresses of tunnel endpoints. These tunnels might be bidirectional or unidirectional. Bidirectional configured tunnels work like a virtual point to point link. Figure 3.4 shows a manually configured tunnel that is utilized to link IPv6 hosts or networks over the IPv4 infrastructure. Generally, these tunnels are utilized when exchanged traffic is regular. Main disadvantage of the technique is that it requires more administration effort when the quantity of tunnels increases [14].

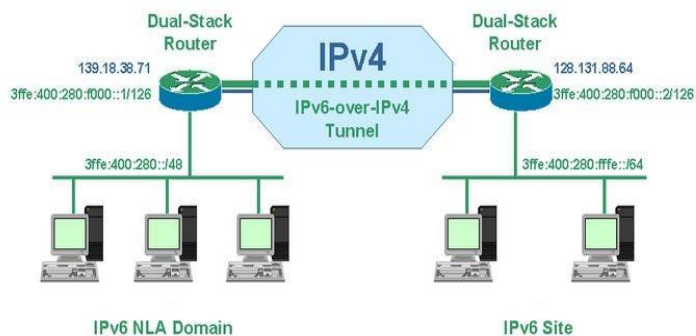

**Figure 2: Configured Tunnel (Manually)**

### 2.5 Automatic Tunneling

If a device directly creates its own tunnels this called 'automatic tunneling'. In this type of tunnel, the address of an IPv4 tunnel endpoint is chosen from the inserted IPv4 address in IPv4 compatible destination address of the tunneled IPv6 packet [16]. So the packet which is being tunneled helps in determining the address of the tunnel endpoint. Figure 3.5 shows Automatic Tunneling technique.

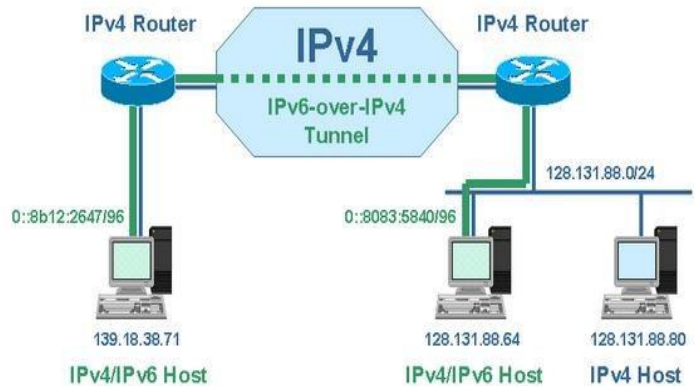

**Figure 4: Automatic Tunneling**

The packet can be delivered through automatic tunneling if the IPv6 address is of type IPv4-compatible address. But when the destination address is IPv6-native, it is not possible to deliver the packet through automatic tunneling. To direct automatic tunneling, a routing table needs to be entered. A specific static routing table entry should be defined for the prefix **0:0:0:0:0:0/96**. Packets matching with this prefix are delivered to a pseudo-interface driver that makes automatic tunneling. Generally, these tunnels can be utilized between individual hosts or networks in which there are incidentally needs for traffic exchanges.

### 2.6 6to4 Automatic Tunneling

In 6 to 4 tunneling technique, tunneling endpoints are configured automatically between devices. 6to4 mechanism means IPv6 traffic is tunneled upon IPv4 networks between separated IPv6 networks. The format of 6to4 network address includes the prefix 2002::/16 after which the universally distinct IPv4 address come [17]. A concatenated form of a 48 prefix is given as an example in Figure 3.6 for the 192.168.99.1 (IPv4 address), 2002:c0a8:6301::/48 prefix of 6to4 address (where c0a8:6301 is the hexadecimal notation for 192.168.99.1).

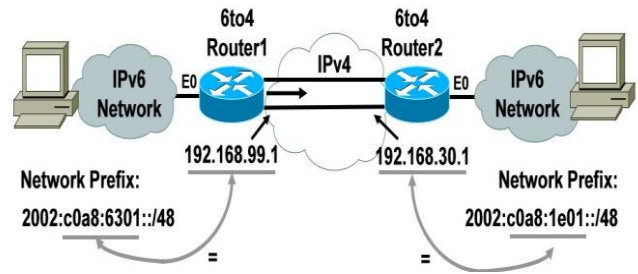

**Figure 5: 6to4 Addresses in a Network [33]**

**2.7    Intra-Site Automatic Addressing Protocol (ISATAP)**

ISATAP is an IPv6 an IPv4 address [43]. The ISATAP tunnel is capable to supply a link between IPv6 and IPv4 routers. When the link is established the host within ISATAP receives an address called local ISATAP address and then host detects the next step of the ISATAP router. The packets are then sent by the tunnel after inserting the IPv6 address into an IPv4 address [35]. At the receiving side, the IPv4 header is deleted and the packet is sent to the IPv6 Host; there the server sends the packets to the ISATAP network and finally the ISATAP router prepares the IPv6 packets into IPv4 and sends them to the ISATAP host, which then removes the IPv4 header and extracts the IPv6 packets transition mechanism used to transmit IPv6 packets between dual-stack nodes on top of an IPv4 network. ISATAP is used to link IPv6 address with the specific prefix fe80::5efe/96 and this address is followed by the IPv4 which is 32 bit as shown in Figure 3.7. The IPv6 address will be within [18].

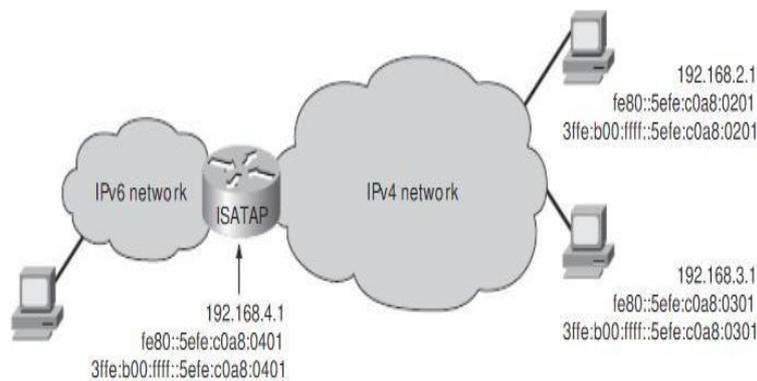

**Figure 6: ISATAP**

**3.0    Translation Mechanism**

Translation mechanism refers to devices capable of direct conversion from the IPv4 protocol to the IPv6 protocol. This mechanism requires translators that can convert IPv4 address to IPv6 address as shown in Figure 3.8. When translation mechanism is used, there is no need for dual-stack network and the network interoperability problem is solved since routers play as communicators. However, translation mechanism faces with limitation problem like IPv4 Network Address Translation (NAT) as well as it is hard to control on larger scale networks.

184

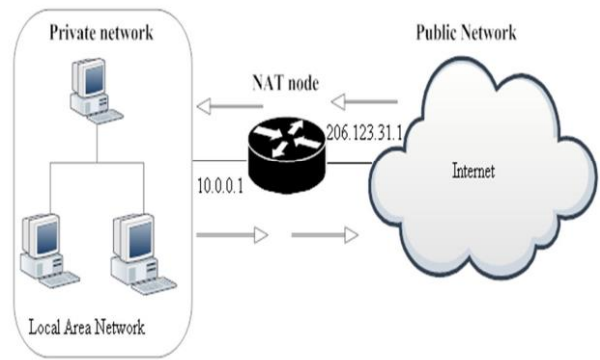

**Figure 7: Translation Mechanisms**

### 3.1 Routing protocols

The selection of a path for transmitting datagrams is called routing. The important task of a router in a network is to determine the best path during the packet forwarding process. The routing process need a router to use routing table and the routing table contains entries information of different paths through the routing protocols. The IPv6 uses the similar kind of routing protocols with IPv4 but with some modifications. However, IPv6 is a new version of protocol and different from IPv4. The routing table is also managed separately from IPv4 routing table when both protocols were enabled on a router.

### 3.2 Exterior Gateway Protocols

Exterior gateways protocols are used to exchange routing information among different Autonomous Systems (AS).

Ø Example of an EGP:-
- Border Gateway Protocol (BGP4+).
- Exterior Gateway Protocol (EGP).

### 1.0 Transition Mechanisms

The deployment of IPv6 is happening gradually on the Internet. Initially deployment of IPv6 should be happened within isolated islands, and then these islands should interconnect with other islands over existing IPv4 Infrastructures. To connect these islands "Transition mechanisms" are needed and these are compatible mechanisms employed for Co-existence of IPv6 and IPv4 infrastructure.

A number of transition mechanisms have been defined to support co-existence [19]. There is an additional need for a node to support IPv6. RFC (request for comments) defines the following types of nodes [19][RFC 4294].

- A node is device that has IP implements in it.
- A router is a device that forwards IP packets towards destination.
- A host is a node but not a router.

**Types of node:**
- **IPv4-only node:** A host or router that implements only IPv4 protocol in it.
- **IPv6/IPv4 node:** A host or router that implements both IPv6 and IPv4 protocol in it.
- **IPv6-only node:** A host or router that implements only IPv6 protocol in it.

The Internet Engineering Task Force (*IETF*) has defined a number of specific mechanisms to assist transition of IPv6 [19]. These mechanisms are basically divided as follows:
➢ Tunneling
➢ Dual Stack

### 1.1 Dual Stack Transition Mechanism (DSTM)

Dual-stack transition mechanism enables to run both IP stacks (IPv4 and IPv6) in a single node. Dual stack nodes maintain both IP protocol stacks that operates parallel and thus allow the end node to use either protocols [39]. The Dual stack node is capable of handling both kinds of IP (IPv4&IPv6) routing. Flow or routing decisions in the node are based on IP header version's field [39]. Both IPv4 and IPv6 shares common transport layer protocols such as TCP/IP. Many of client and server operating systems provide dual IP protocol stacks [19].

For example: Windows XP, Vista, 7, Windows server 2003, Linux, Mac OS x [10][19]. TCP/IP model for dual stack node is as follows:

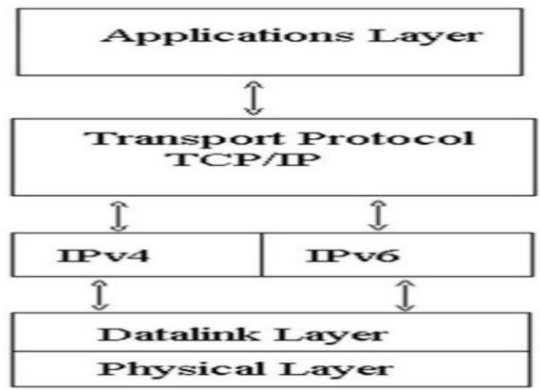

**Figure 8: Dual stack TCP/IP model**

Dual stack networking deploys IPv4 and IPv6 in the same infrastructure. If a node that support dual stack network, should be able to understand and process both IP protocols network. The dual stack node itself cannot decide at randomly, which IP stacks to use to communicate so the routing protocol decides, which stack to use. Example of Dual Stack infrastructure as follows:

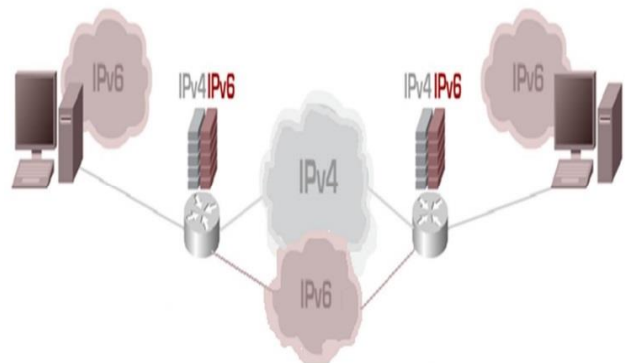

**Figure 9: Dual Stack Infrastructure**

Some of them are as follows:
Static Tunneling [19]
- Automatic Tunneling using IPv4-compatible Addresses [19]
- 6over4 Transition Mechanism [19]
- 6to4 Transition Mechanism [19]
- Intrasite Automatic Tunnel Addressing protocol (ISATAP)[19].

Example of a tunneling infrastructure:

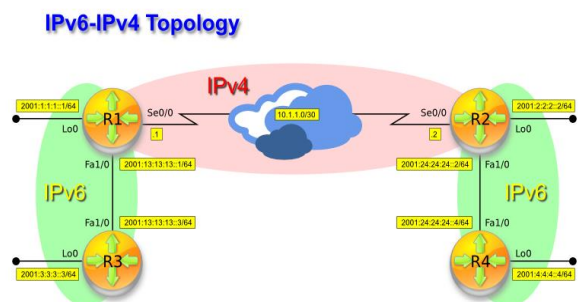

**Figure 10: Tunneling Mechanism infrastructure**

### 5.0 Implementation

The implementation setup has been done between "headquarters" and "branch office" of a company over public network (Internet Service Provider). Two sample models were experimented in the simulator environment to evaluate the complexity and pros and cons of each method. Implementation work is done in two scenarios by implementing two methods such as 6to4 tunnel and Dual stack.

- **Scenario 1 method: 6to4 Tunneling.**
- **Scenario 2 method: Dual stack.**

Behind selection of these particular two methods are easy to implement in existed equipment in an organization instead of spending budget on new equipment and accessories. Basic topology was established with three routers in which we have three server named as headquarters (HQ), Internet service provider (ISP 1) and Branch office (Br). Detailed process of connectivity has been explained in each of the scenarios. In both scenarios the connectivity was same.

**Equipment used:-** **Routers:** Cisco 2811 series with Cisco IOS version 12.4(4) T8.
-   **Client:** Windows 7 with IP dual stack installed.

**Scenario 1 (6to4 tunnel)**

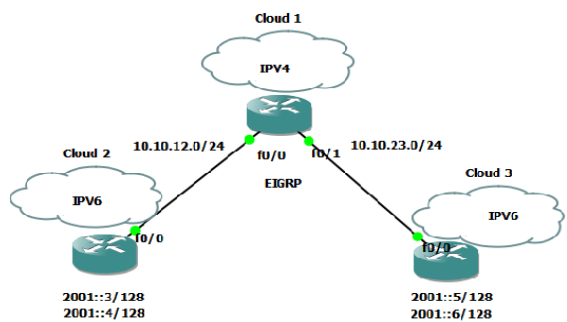

**Figure 11: 6to4 tunnel network topology**

**5.1 Physical Connections:**

A network has been established between three routers headquarters (HQ)-R1 and branch office (Br) R3 over the Internet service provider (ISP1)-R2 as shown in the figure 4.5.

In Scenario 1 to setup a network, the three routers (R1, R2, R3) were connected to each other. From the figure 4.5. Router-1 (HQ) is connected to with interface fa0/0 to Router-2 (ISP). Router-2 with interface fa0/1 was connected to Router-2 (Br). This fulfill physical connectivity between headquarter to branch office.

**IP Address Scheme:**
**Headquarter:**

**Table 1: Headquater IP addresses**

| Interface | IPV4 Address | IPV6 Address |
|---|---|---|
| Fast Ethernet 0/0 | 10.10.12.1 | ----- |
| Loopback 0 | ------- | 2001::3/128 |
| Loopback 0 | ------- | 2001::4/128 |
| Tunnel 0 | -------- | 2001::7/128 |

*Table 1: ISP 1 IP addresses*
**5.2 Establish routing**:

To make communication possible from Router1 to Router2, routing protocols must be established in all routers. To achieve these, two kinds of protocols were employed in a network. Those are, BGP as an Exterior gateway protocol (EGP) for public network such as Internet service provider network and Interior gateway protocol (IGP) is OSPFv3 for private network such as local connections. In particular selection of these protocols is, OSPFv3 is a link state protocol makes faster network routes merging and maintains copies of routing tables especially supports IPv6 routing. BGP is a path vector routing protocol, present widely using as EGP protocol in the Internet. Two IP protocols were employed in this scenario such as IPv4 for public networks and IPv6 for private networks. Public network was used between HQ to ISP1 and ISP1 to Br in **Figure 4.5**. IP addresses used for communication were in this scenario explained in above exercise.

Private network was used to connect between R1 to R2 & R2 to R3. In router HQ, OSPFv3 configured for IPv6 network and BGP for IPv4 network. Router ISP1 is in public network so BGP was configured. In router Br, OSPFv3 configured for IPv6 network and BGP for IPv4 network. Routing protocols were established in all routers but the incompatibility of IPv4 and IPv6 does not make communication possible from Host1 to Host2. So in this situation a smooth transition mechanism is needed to make this communication possible.

Well, **what mechanism can make this possible and how?**

As explained in above 6to4 manual tunnel is one of the tunneling mechanisms. It used for encapsulate IPv6 packets into IPv4 packets in a dual stack enabled router and send over normal IPv4 routing to the end of the tunnel, node at the tunnel end is also a dual stack enabled router decapsulate IPv6 packets from IPv4 packets and delivered to according destination.

**6to4 manual tunnel Process:**
In Scenario1 an IPv6 packet from Router 1 generated to the destination as Router2, and sent to ISP. Router HQ is a tunnel starting point that encapsulate IPv6 packet in IPv4, and sent through ISP1 over normal IPv4 routing to the end of tunnel. End of the tunnel is a router BR decapsulate the IPv6 packet from the IPv4 packet and delivered to Router2.

### 3.3 Scenario 2 (Dual Stack)

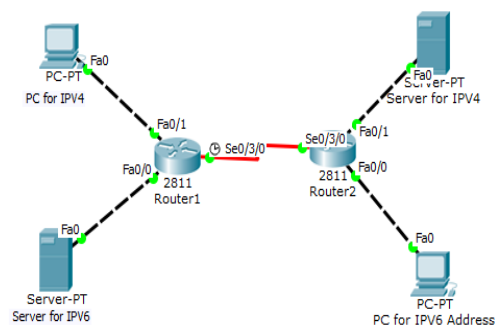

**Figure 12: Dual stack network topology**

**CONCLUSIONS**

In this thesis, we examine the present transitioning techniques from IPv4 to IPv6. We have also done some configurations on the generally utilized transition mechanisms like Automatic 6to4 tunneling and Dual-Stack. We have compared the transition mechanisms with each other between IPv6 and IPv4.

IPv6 larger address space provides more unique globally unicast addresses for the present and future Internet growth. Fully deployment of IPv6 needs upgrading of applications, hosts, routers and DNS to support IPv6, might be expensive and deployment takes many years. These situation transition mechanisms are one of the best solutions, so this makes IPv6 & IPv4 networks run in the same infrastructure. IPv4 to IPv6 several transition mechanisms have been developed for according to different organization needs.